\documentclass[aps,prl,twocolumn,floatfix,superscriptaddress]{revtex4}

\usepackage{amssymb,amsmath,amstext}                
\usepackage{graphicx}
\usepackage{epstopdf}                                               
\usepackage{color}                                                     
\usepackage{bm}                                                        
\usepackage{appendix}

\usepackage[utf8]{inputenc}
\usepackage{lipsum}

\usepackage{ulem}
\usepackage{latexsym}
\usepackage[colorlinks=true,citecolor=blue,linkcolor=magenta]{hyperref}
\usepackage{soul}
\usepackage{amsmath}
\usepackage{float}

\DeclareMathOperator{\sinc}{sinc}

\newcommand{\vect}[1]{\mathbf{#1}}

\def\be{\begin{equation}}
\def\ee{\end{equation}}
\def\bea{\begin{eqnarray}}
\def\eea{\end{eqnarray}}
\def\ra{\rangle}
\def\la{\langle}
\def\bi{\begin{itemize}}
\def\ei{\end{itemize}}

\definecolor{dgreen} {RGB}{78,138,21}

\definecolor{xuedong} {RGB}{0,128,0}

\definecolor{giergiel} {RGB}{0,168,128}

\definecolor{purple} {RGB}{128,0,160}

\begin{document}
\title{
Topological Molecules and Topological Localization of a Rydberg Electron on a Classical Orbit}
\author{Ali Emami Kopaei}
\thanks{These two authors contributed equally}
\affiliation{
Instytut Fizyki Teoretycznej,
Uniwersytet Jagiello\'nski, ulica Profesora Stanis\l{}awa \L{}ojasiewicza 11, PL-30-348 Krak\'ow, Poland}
\author{Xuedong Tian}
\thanks{These two authors contributed equally}
\affiliation{
Instytut Fizyki Teoretycznej,
Uniwersytet Jagiello\'nski, ulica Profesora Stanis\l{}awa \L{}ojasiewicza 11, PL-30-348 Krak\'ow, Poland}
\affiliation{College of Physics Science and Technology, Guangxi Normal University, 541004, Guilin, China}
\author{Krzysztof Giergiel}
\affiliation{
Instytut Fizyki Teoretycznej,
Uniwersytet Jagiello\'nski, ulica Profesora Stanis\l{}awa \L{}ojasiewicza 11, PL-30-348 Krak\'ow, Poland}
\affiliation{Optical Sciences Centre, Swinburne University of Technology, Melbourne, Australia}
\author{Krzysztof Sacha}
\affiliation{
Instytut Fizyki Teoretycznej,
Uniwersytet Jagiello\'nski, ulica Profesora Stanis\l{}awa \L{}ojasiewicza 11, PL-30-348 Krak\'ow, Poland}

\begin{abstract}
It is common knowledge that atoms can form molecules if they attract each other. Here, we show that it is possible to create molecules where bound states of the atoms are not the result of attractive interactions but have the topological origin. That is, the bound states of the atoms correspond to the topologically protected edge states of a topological model. Such topological molecules can be realized if the interaction strength between ultra-cold atoms is properly modulated in time. A similar mechanism allows one to realize topologically protected localization of an electron on a classical orbit if a Rydberg atom is perturbed by a properly modulated microwave field.
\end{abstract}
\date{\today}

\maketitle

Experiments often suffer from imperfections and external perturbations that are difficult to eliminate. The situation can change if the state which has to be realized in the laboratory is protected by topology. The phenomena determined by topological invariants are robust unless a perturbation is so strong that it changes the topology. For example, we may dramatically deform a torus, but it is still the same topological object unless we cut it. The range of topologically protected phenomena is broad, from the quantum Hall effect to ideas of topological quantum computation \cite{NayakRMP2008,Hasan2010}.

In this Letter, we address the question of whether bound states of atoms or localization of a Rydberg electron on a classical orbit can be protected by topology. It would be vital to experimentalists because these objects would be robust and resistant to external perturbations. To accomplish our objectives, we employ {\it time engineering} developed in the field of time crystals and phase space crystals, which allows realization of condensed matter physics in the time domain \cite{Guo2013,Sacha2015,Sacha15a} (for reviews, see \cite{Sacha2017rev,guo2020,SachaTC2020,GuoBook2021}). Basically it relies on resonant periodic perturbation of, e.g., ultra-cold atoms, which effectively behave like a solid-state system.

{\bf Localization of a Rydberg electron.}
Let us begin with a highly excited hydrogen atom. Even if the electron is highly excited and prepared in a localized wavepacket, the classical picture of a particle moving on a Kepler orbit is quickly lost due to the spreading of the wavepacket over the entire classical orbit \cite{Buchleitner2002}. Spreading can be suppressed if the atom is driven resonantly by an electromagnetic field \cite{Delande1994,Bialynicki1994,Buchleitner2002}. Then, in the frame moving along a Kepler orbit, a potential well is created that supports the wavepacket, and the classical picture of the electron circulating around the nucleus can be experimentally demonstrated \cite{Maeda2004,Maeda2007,Maeda2009,Wyker2012}. There is also an idea to prevent the electron from spreading by employing a fluctuating microwave field \cite{Giergiel2017}. That is, a fluctuating microwave field induces destructive interference phenomena that are responsible for Anderson localization of an electronic wavepacket on a classical orbit. Signatures of Anderson localization of a Rydberg electron can also be observed if ground-state atoms are immersed within a Rydberg wavefunction \cite{Eiles2021}.

To show that the electron in a hydrogen atom can be represented by a wavepacket whose localization is protected by topology, let us consider a H atom in the presence of a static electric field and two linearly polarized microwave fields with frequencies $\omega$ and $2\omega$ and a certain superposition of their subharmonics and harmonics $f(t)=f(t+s2\pi/\omega)=\sum_{k\ne 0}f_ke^{ik\omega t/s}$ where $s$ is integer. The Hamiltonian of the system in the atomic units reads
\be
H=\frac{\vect{p}^2}{2}-\frac{1}{r}+z[F+F_1\cos(\omega t)+F_2\cos(2\omega t)+\lambda f(t)],
\label{h}
\ee
where $F$, $F_{j}$ and $\lambda$ are the strength of the electric field and the amplitudes of the microwave fields, respectively. Before we present quantum results, we derive a classical effective Hamiltonian that allows us to understand how the phenomenon, in which we are interested, emerges.

The natural variables for the classical description of an H atom are the action angle variables, where $I$ is the principal action (the classical analogue of the principal quantum number $n$), $L$ is the total angular momentum, and $\theta$ and $\psi$ are the corresponding conjugate canonical variables that describe the position of the electron in a Kepler orbit and the angle between the major axis of the elliptical orbit and the $z$ axis, respectively \cite{Buchleitner2002}. We assume that the projection of the angular momentum of the electron on the $z$ axis, which is a constant of motion, is zero. We also assume that the microwaves are resonant with unperturbed electronic motion, i.e. the ratio of the frequency $\omega$ and the frequency of the electronic motion is an integer number $s=\omega I_s^3$ where $I_s$ is the resonant value of the principal action [$s$ is the same as in the expression of $f(t)$]. In the frame moving along the orbit, $\Theta=\theta-\omega t/s$, all dynamical variables are slowly varying if we are close to the resonant trajectory, i.e., if $P=I-I_s\approx 0$. By averaging the Hamiltonian over time $t$, we obtain the effective (secular) Hamiltonian \cite{Lichtenberg1992,SM},
\be
H_{\rm eff}=\frac{P^2}{2m_{\rm eff}}+V_1\cos(s\Theta)+V_2\cos(2s\Theta)+\lambda V_{\rm b}(\Theta),
\label{heff}
\ee
where a constant term is omitted, the effective mass $m_{\rm eff}=-I_s^4/3$, $V_{1}=-I_{s}^{2}F_{1}\mathcal{J}_{s}^{\prime}(s)/s$, $V_{2}%
=-I_{s}^{2}F_{2}\mathcal{J}_{2s}^{\prime}(2s)/2s$ and $V_{\rm b}(\Theta)=-I_{s}^{2}\sum_{k\neq0}f_{-k}%
\mathcal{J}_{k}^{\prime}(k)e^{ik\Theta}/k$,  where ${\cal J}_k'$'s are derivatives of the ordinary Bessel functions. The effective Hamiltonian (\ref{heff}) describes the motion of the electron around the elongated Kepler ellipse with the total angular mometum $L=0$. This orbit is aligned along the $z$ axis ($\psi=0$) and is stable, provided that the static electric field is sufficiently strong \cite{Sacha1998,SM}.

\begin{figure}[t]
\includegraphics[width=.95\columnwidth]{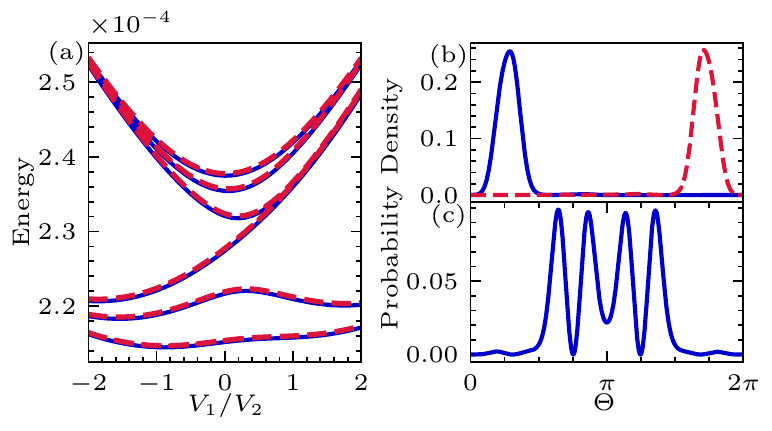}
\caption{Topological localization of a Rydberg electron. Panel~(a) shows the spectrum of the quantized version of the classical Hamiltonian (\ref{heff}) (red dashed lines) and the corresponding energy levels of the quantum effective Hamiltonian (\ref{qheff}) (solid blue lines) for $s=4$. For $V_1/V_2>0$, two edge states of the Hamiltonian (\ref{heff}) form with nearly degenerate energies visible in the middle of the spectrum. Energy values are presented in the atomic units multiplied by $n_s^2$ and the constant term $-3/2$ is subtracted. Panel~(b) presents probability densities of these two edge states for $V_1/V_2=2$ --- the edge is located around $\Theta=0$ (or equivalently $\Theta=2\pi$) and the edge states localize close to it. Other eigenstates (bulk states) are delocalized along the entire range of $\Theta$ [in (c) the fifth excited eigenstate of (\ref{heff}) is shown]. The parameters of the system are the following: the resonant principal quantum number $n_s=I_s=800$, $n_s^3\omega=4$, $n_s^4F=1.5\times10^{-4}$, $n_s^4F_1=1.258\times10^{-3}$, $n_s^4\lambda=1.172\times10^{-5}$, and $f_k=e^{ik\epsilon}\cos(k\pi/21)\sinc^{2}(k\pi/14)k/J_{k}^{\prime}(k)$ with $\epsilon=5\times10^{-3}$ for $|k|\le 20$ and $f_k=0$ for higher $|k|$ \cite{foot1}. In (b) and (c), $n_s^4F_2=1.93\times10^{-3}$.}
\label{fig1} 
\end{figure}

If $\lambda=0$ and $s\gg 1$, the Hamiltonian (\ref{heff}) shows that the electron in an H atom, in the frame moving along the elongated resonant orbit, behaves the same way as an electron in a crystalline structure with a two-point basis. Actually, if we quantize the classical Hamiltonian (\ref{heff}) and restrict ourselves to its first two energy bands, we arrive at the SSH model \cite{Su1979}, which is topologically non-trivial if $V_1/V_2>0$, i.e., it is characterized by the non-zero winding number \cite{Asboth2016short,Cooper2019}. This is a simple model of a topological insulator that has been investigated theoretically and experimentally in many different systems \cite{Atala2013,Meier2016,St-Jean2017,Giergiel2018b,Asboth2016short}. Striking property of topological insulators with edges is the presence of topologically protected states that are localized at the edges \cite{Asboth2016short}. To introduce an edge in the potential in (\ref{heff}), we turn on the subharmonics and harmonics of the microwave field ($\lambda\ne 0$) prepared so that the additional potential $\lambda V_{\rm b}(\Theta)$ describes a barrier located at a certain $\Theta$. Then, the spectrum of the quantized version of $H_{\rm eff}$ reveals a pair of energy levels in the gap between two bands and the corresponding eigenstates are localized close to the edge; see Fig.~\ref{fig1}.

\begin{figure}[t]
\includegraphics[width=.95\columnwidth]{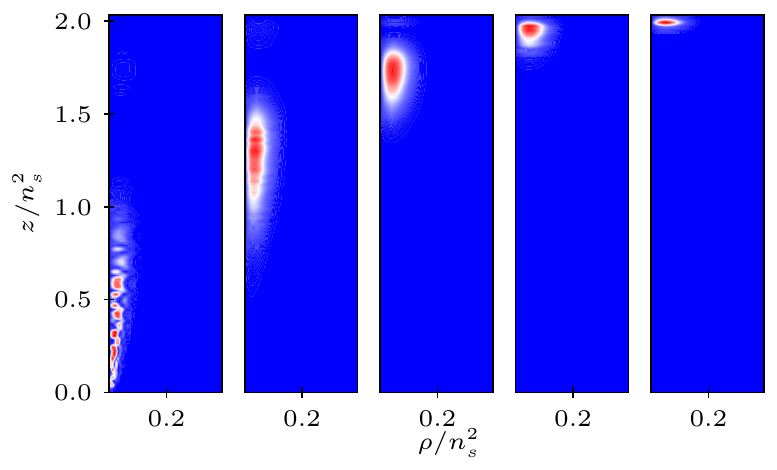}
\caption{Topological localization of a Rydberg electron.
Probability densities of an eigenstate of (\ref{qheff}) corresponding to one of the edge states  identified with the help of (\ref{heff}). The densities are plotted in the laboratory frame and in the cylindrical coordinate for different moments of time, i.e. $\omega t/s=7\pi/4+j\pi/4$ where integer $j$ goes from 0 to 4 from left to right panels, respectively. The presented edge state reveals the wavepacket evolving on the elongated Kepler orbit  with the period $s2\pi/\omega$ and its localization is protected by topology. The sequence of the panels illustrates time evolution of the wavepacket duirng one half of the period. During the other half, the time evolution can be illusrated by the same panels but ordered from right to left.  The parameters are the same as in Fig.~\ref{fig1}(b).}
\label{fig2}
\end{figure}

So far we have described a H atom by means of the classical effective Hamiltonian which at the end is quantized. However, the entire description can be performed fully quantum mechanically starting with (\ref{h}). When we switch to the moving frame by means of the unitary transformation $e^{i\hat n \omega t/s}$, where $\hat n$ is the principal quantum number operator, and average the Hamiltonian over $t$, we obtain the matrix of the quantum effective Hamiltonian
\bea
\la n',l'|\hat H_{\rm eff}|n,l\ra &=&\left(-\frac{1}{2n^2}-n\frac{\omega}{s}\right)\delta_{nn'}\delta_{ll'}+\la n',l'|z|n,l\ra
\cr && \times \left[F\delta_{nn'}+F_1(\delta_{n+s,n'}+\delta_{n-s,n'})/2\right.
\cr &&
\left. +F_2(\delta_{n+2s,n'}+\delta_{n-2s,n'})/2
+\lambda f_{n-n'}\right],
\cr &&
\label{qheff}
\eea
where $|n,l\ra$ is a hydrogenic eigenstate with the principal quantum number $n$, total angular momentum $l$, and the projection of the angular momentum on the z axis equal to zero. The spectra of the quantized classical Hamiltonian (\ref{heff}) and the quantum Hamiltonian (\ref{qheff}) are compared in Fig.~\ref{fig1}. In Fig.~\ref{fig2} we present the time evolution, in the laboratory frame, of an eigenstate of (\ref{qheff}) corresponding to one of the edge states identified with the help of the Hamiltonian (\ref{heff}). One can see that the electron is represented by a wavepacket that periodically propagates on the elongated ellipse and does not spread because its localization is protected by topology. This behavior is resistant to a perturbation of the effective Hamiltonian, unless the perturbation is so strong that the gap between the bands is closed \cite{SM}.

In Ref.~\cite{Buchleitner2002} non-spreading wavepackets of a Rydberg electron are described. However, their formation has nothing to do with topological protection. They are related to the trapping of a Rydberg electron in a single well of an effective potential that is created by the $1:1$ resonant driving of the H atom. The mechanism of the formation of the localized states considered in this Letter is completely different. The localized states are edge states with localization length $\xi\propto 1/\ln(J'/J)$, where $J$ and $J'$ are tunneling rates of an electron through the higher and lower barriers, respectively, of the effective potential in (\ref{heff}) \cite{Asboth2016short}. The localization length $\xi$ can be much larger than the size of a single well of the effective potential in (\ref{heff}). Topological protection means that if we change the parameters of the effective model (\ref{heff}), $\xi$ can change but the localization phenomenon itself will not be broken, provided that the energy gap is not closed.

{\bf Topological molecules.}
Attractive interactions between charge particles allow atoms to form molecules. It is also possible to create bound states of atoms if the interaction potential between them changes in a disordered way as a function of their relative distances. Then, they can form the so-called Anderson molecules that are created due to destructive interference and the resulting Anderson localization \cite{Giergiel2018,SachaTC2020,Matus2021}. Here we show that it is possible to realize bound states of atoms which have a topological origin. We will see that the description of two atoms with the interaction strength modulated in time can be reduced to the effective Hamiltonian where we can identify a {\it center of mass} degree of freedom described by the free particle-like Hamiltonian and the {\it relative position} degree of freedom described by the Hamiltonian (\ref{heff}). The edge states of the Hamitlonian (\ref{heff}) will correspond to the molecular bound states protected by topology.

Let us consider two atoms which are moving in the one-dimensional (1D) infinite well potential. For simplicity, we assume that they are the same atomic species but in different hyperfine states. The infinite well potential in 1D can be created experimentally if, in a 3D trap, with strong transverse confinement, two barriers are implemented that limit the motion of the atoms along the longitudinal direction \cite{Gaunt2013}. At low kinetic energies, the interaction between atoms can be modeled by means of the contact Dirac delta potential with the strength $g$ proportional to the s-wave scattering length of the atoms. We assume that $g(t)$ is modulated periodically in time by means of Feshbach resonance \cite{Chin2010} or confinement-induced resonance \cite{Olshanii1998} so that the Hamiltonian of the system reads
\be
H=\frac{p_1^2+p_2^2}{2}+g(t)\delta(x_1-x_2),
\label{hm}
\ee
where $g(t)=F_1\cos(\omega t)+F_2\cos(2\omega t)+\lambda f(t)$. We have used $\pi^2\hbar^2/mR^2$ and $R/\pi$ as units of energy and length, respectively, where $R$ is the size of the potential well and $m$ is the mass of the atoms.
For simplicity, we have used the same notation for the parameters of $g(t)$ as in the case of a H atom, cf.~(\ref{h}), but now $F_j$ and $\lambda$ are quantities proportional to the s-wave scattering length of the atoms. As previously, $f(t)=f(t+s2\pi/\omega)=\sum_{k\ne 0}f_{k}e^{ik\omega t/s}$. In the units that we use, the motion of the atoms is limited between $0$ and $\pi$ due to the presence of the potential walls. Thus, the wavefunction vanishes if $x_1$ or $x_2$ equals 0 or $\pi$.

To describe the system, it is sufficient to restrict to the subspace of symmetric combinations of the eigenstates of the noninteracting atoms, i.e. $\Phi_{n_1,n_2}=[\phi_{n_1,n_2}+\phi_{n_2,n_1}]/\sqrt{2}$ for $n_1>n_2$ and $\Phi_{n_1,n_1}=\phi_{n_1,n_1}$, where $\phi_{n_1,n_2}(x_1,x_2)=2\sin(n_1x_1)\sin(n_2 x_2)/\pi$, because there is no interaction between the atoms in the antisymmetric subspace \cite{SM}. We are interested in the resonant behavior of the system where $n_1\approx \omega/2s$ and $n_2\approx \omega/2s$. Switching to the moving frame by means of the unitary transformation $e^{i(\hat n_1+\hat n_2)\omega t/2s}$ and averaging the resulting Hamiltonian over time, we arrive at the matrix of the quantum effective Hamiltonian $\hat H_{\rm eff}$ that consists of independent blocks labeled with different values of $n_{\rm cm}=n_1-n_2$. That is, in a given $n_{\rm cm}$ block the matrix elements of the $\hat H_{\rm eff}$ read \cite{SM}
\bea
\la n'|\hat H_{\rm eff}|n\ra
&=&n^2\delta_{n',n}+\frac{F_1}{2\pi}(\delta_{n',n+s}+\delta_{n',n-s})
\cr &&
+\frac{F_2}{2\pi}(\delta_{n',n+2s}+\delta_{n',n-2s})+\frac{\lambda}{\pi}f_{n'-n}
\cr &&
+\frac{n_{\rm cm}^2}{4}\delta_{n',n},
\label{mqheff}
\eea
where $n=(n_1+n_2)/2-\omega/2s$. The position operators conjugate to the quantum numbers $n_{\rm cm}$ and $n$ are $x_{\rm cm}=(x_1-x_2)/2$ and $x=x_1+x_2$, respectively. This effective description is valid in the subspace $|n|\ll \omega/2s$.

Apart from the last term, the matrix (\ref{mqheff}) is identical to the matrix of the quantized version of the classical effective Hamiltonian (\ref{heff}) calculated in the plane wave basis, $e^{in\Theta}/\sqrt{2\pi}$, if the effective mass $m_{\rm eff}=1/2$, $V_j=F_j/\pi$ and $V_b=\sum_{k\ne0}f_{k}e^{ik\Theta}/\pi$. Thus, the two-atom system behaves like a two-particle system whose center of mass position $x_{\rm cm}$ (conjugate to the momentum quantum number $n_{\rm cm}$) moves like a free particle [cf. the term $n_{\rm cm}^2/4$ in (\ref{mqheff})] while the relative position $x$ can possess topologically protected edge states corresponding to the bound states of the particles. Particles prepared in an edge state form a topological molecule in which their relative position $x$ is described by a localized state that is protected by topology.

Our secular approximation approach allows us to identify parameters suitable for the realization of a topological molecule, but we are also able to describe the system exactly by numerical diagonalization of the Floquet Hamiltonian $H_F=H-i\partial_t$ where $H$ is given in (\ref{hm}) \cite{Shirley1965,Buchleitner2002,SM}. The eigenstates of $H_F$ are called Floquet states which evolve with the same period as the period of $g(t)$ and the corresponding eigenvalues are quasienergies of the system. The eigenstates of the effective Hamiltonian (\ref{mqheff}) also evolve periodically in time when we return to the laboratory frame, and they constitute a very good approximation of the exact Floquet eigenstates of $H_F$.

\begin{figure}[t!]
\includegraphics[width=.95\columnwidth]{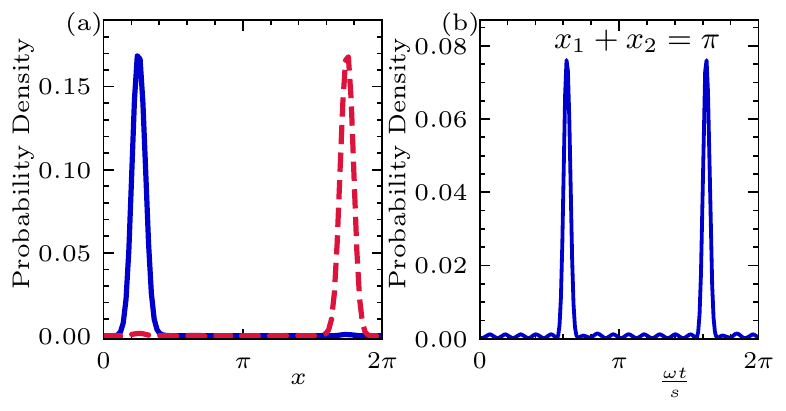}
\caption{Topological molecules. Panel (a) shows probability densities of two edge states in the moving frame. The densities are localized around the certian position $x$, indicating that two atoms form bound states. Panel (b) presents signatures of the formation of the topological molecule in the laboratory frame. If the atoms are prepared in, e.g., the edge state drwan with the blue line in (a), the probability density for the measurement of the atoms at $x_1+x_2=\pi$ reveals a periodic appearance of the localized edge state. The results are obtained within the exact diagonalization of the Floquet Hamiltonian, but they are indistinguishable from the densities obtained with the help of the effective Hamiltonian (\ref{mqheff}). The parameters of the system are the following: $F_1/F_2=2$ with $F_2=10$, $\omega=10^5$, $\lambda=35$ and the Fourier components of $f(t)$ are $f_k=e^{ik \epsilon}\cos(k\pi/21)\sinc^{2}(k\pi/14)$ with $\epsilon=10^{-4}$ for $|k|\le 40$ and $f_k=0$ for greater $|k|$.}
\label{fig3}
\end{figure}

In Fig.~\ref{fig3}(a) we show the exact Floquet states, in the moving frame, that correspond to the edge states described above. The probability densities describing the position $x$, which is conjugate to the quantum number $n$, perfectly match the edge state probability densities obtained with the help of the effective Hamiltonian (\ref{mqheff}). It confirms that the atoms can form topologically protected bound states. When the system is observed in the laboratory frame, the signatures of the formation of the topological molecule are illustrated in Fig.~\ref{fig3}(b). That is, the probability density for the atoms to be observed at $x_1+x_2=\pi$ reveals a periodic appearance of the localized edge state.

{\bf Conclusions.}
We have shown that topology is capable of protecting the states of the atomic constituents. A Rydberg electron can be represented by a localized wavepacket moving along a classical orbit, and such a localization is protected by topology. A similar mechanism is responsible for the formation of topological molecules --- bound states of particles protected by topology.

We have considered the localization of a Rydberg electron on a Kepler orbit aligned along the static electric field direction. However, one may expect that the topological localization can also be observed on Kepler orbits with different shapes. That is, the static electric field allows one to control the shape of a stable Kepler ellipse \cite{Sacha1998}, and similar microwave fields as we have used here should lead to the topological localization of the electron on the ellipse. The experimental setup for realization of the phenomenon we predict can be very similar to the procedure used in Refs.~\cite{Maeda2004,Maeda2007,Maeda2009,Wyker2012} where nonspreading Rydberg wavepackets were demonstrated.

Ultra-cold atoms in a quasi-1D squared well potential \cite{Gaunt2013} seem to be a suitable system for the realization of topological molecules. Two localized clouds of atoms can be pushed to motion, and if it is done at proper phase of the time periodic modulation of the strength of interactions between atoms, the clouds will be locked to a localized edge state of an effective topological model that describes the system \cite{SM}. Modulation of the interaction strength can be done by means of the Feshbach resonance \cite{Chin2010} or the confinement-induced resonance in quasi--1D potentials \cite{Olshanii1998,Haller2010}.

{\bf Acknowlegement.}
This work was supported by the National Science Centre, Poland, through Projects: No. 2018/31/B/ST2/00349 (AEK), No. 2016/20/W/ST4/00314 and No. 2019/32/T/ST2/00413 (KG), QuantERA Programme No. 2017/25/Z/ST2/03027 (KS). KG acknowledges the support of the Foundation for Polish Science (FNP) and the support of the Polish National Agency for Academic Exchange Bekker Programme (BPN/BEK/2021/1/00339). XT acknowledges the support of the Foundation (Yucai
project) from Guangxi Normal University. Some parts of our numerical schemes are based on the PETSc~\cite{petsc-user-ref,petsc-efficient} and SLEPc~\cite{Hernandez:2005} libraries. Numerical computations in this work were supported in part by PL-Grid Infrastructure.


\begin{thebibliography}{39}
\expandafter\ifx\csname natexlab\endcsname\relax\def\natexlab#1{#1}\fi
\expandafter\ifx\csname bibnamefont\endcsname\relax
  \def\bibnamefont#1{#1}\fi
\expandafter\ifx\csname bibfnamefont\endcsname\relax
  \def\bibfnamefont#1{#1}\fi
\expandafter\ifx\csname citenamefont\endcsname\relax
  \def\citenamefont#1{#1}\fi
\expandafter\ifx\csname url\endcsname\relax
  \def\url#1{\texttt{#1}}\fi
\expandafter\ifx\csname urlprefix\endcsname\relax\def\urlprefix{URL }\fi
\providecommand{\bibinfo}[2]{#2}
\providecommand{\eprint}[2][]{\url{#2}}

\bibitem[{\citenamefont{Nayak et~al.}(2008)\citenamefont{Nayak, Simon, Stern,
  Freedman, and Das~Sarma}}]{NayakRMP2008}
\bibinfo{author}{\bibfnamefont{C.}~\bibnamefont{Nayak}},
  \bibinfo{author}{\bibfnamefont{S.~H.} \bibnamefont{Simon}},
  \bibinfo{author}{\bibfnamefont{A.}~\bibnamefont{Stern}},
  \bibinfo{author}{\bibfnamefont{M.}~\bibnamefont{Freedman}}, \bibnamefont{and}
  \bibinfo{author}{\bibfnamefont{S.}~\bibnamefont{Das~Sarma}},
  \bibinfo{journal}{Rev. Mod. Phys.} \textbf{\bibinfo{volume}{80}},
  \bibinfo{pages}{1083} (\bibinfo{year}{2008}),
  \urlprefix\url{https://link.aps.org/doi/10.1103/RevModPhys.80.1083}.

\bibitem[{\citenamefont{Hasan and Kane}(2010)}]{Hasan2010}
\bibinfo{author}{\bibfnamefont{M.~Z.} \bibnamefont{Hasan}} \bibnamefont{and}
  \bibinfo{author}{\bibfnamefont{C.~L.} \bibnamefont{Kane}},
  \bibinfo{journal}{Rev. Mod. Phys.} \textbf{\bibinfo{volume}{82}},
  \bibinfo{pages}{3045} (\bibinfo{year}{2010}),
  \urlprefix\url{https://link.aps.org/doi/10.1103/RevModPhys.82.3045}.

\bibitem[{\citenamefont{Guo et~al.}(2013)\citenamefont{Guo, Marthaler, and
  Sch\"on}}]{Guo2013}
\bibinfo{author}{\bibfnamefont{L.}~\bibnamefont{Guo}},
  \bibinfo{author}{\bibfnamefont{M.}~\bibnamefont{Marthaler}},
  \bibnamefont{and} \bibinfo{author}{\bibfnamefont{G.}~\bibnamefont{Sch\"on}},
  \bibinfo{journal}{Phys. Rev. Lett.} \textbf{\bibinfo{volume}{111}},
  \bibinfo{pages}{205303} (\bibinfo{year}{2013}),
  \urlprefix\url{https://link.aps.org/doi/10.1103/PhysRevLett.111.205303}.

\bibitem[{\citenamefont{Sacha}(2015{\natexlab{a}})}]{Sacha2015}
\bibinfo{author}{\bibfnamefont{K.}~\bibnamefont{Sacha}},
  \bibinfo{journal}{Phys. Rev. A} \textbf{\bibinfo{volume}{91}},
  \bibinfo{pages}{033617} (\bibinfo{year}{2015}{\natexlab{a}}),
  \urlprefix\url{http://link.aps.org/doi/10.1103/PhysRevA.91.033617}.

\bibitem[{\citenamefont{Sacha}(2015{\natexlab{b}})}]{Sacha15a}
\bibinfo{author}{\bibfnamefont{K.}~\bibnamefont{Sacha}}, \bibinfo{journal}{Sci.
  Rep.} \textbf{\bibinfo{volume}{5}}, \bibinfo{pages}{10787}
  (\bibinfo{year}{2015}{\natexlab{b}}),
  \urlprefix\url{https://www.nature.com/articles/srep10787}.

\bibitem[{\citenamefont{{Sacha} and {Zakrzewski}}(2018)}]{Sacha2017rev}
\bibinfo{author}{\bibfnamefont{K.}~\bibnamefont{{Sacha}}} \bibnamefont{and}
  \bibinfo{author}{\bibfnamefont{J.}~\bibnamefont{{Zakrzewski}}},
  \bibinfo{journal}{Rep. Prog. Phys.} \textbf{\bibinfo{volume}{81}},
  \bibinfo{pages}{016401} (\bibinfo{year}{2018}),
  \urlprefix\url{https://doi.org/10.1088/1361-6633/aa8b38}.

\bibitem[{\citenamefont{Guo and Liang}(2020)}]{guo2020}
\bibinfo{author}{\bibfnamefont{L.}~\bibnamefont{Guo}} \bibnamefont{and}
  \bibinfo{author}{\bibfnamefont{P.}~\bibnamefont{Liang}},
  \bibinfo{journal}{New Journal of Physics} \textbf{\bibinfo{volume}{22}},
  \bibinfo{pages}{075003} (\bibinfo{year}{2020}),
  \urlprefix\url{https://doi.org/10.1088/1367-2630/ab9d54}.

\bibitem[{\citenamefont{Sacha}(2020)}]{SachaTC2020}
\bibinfo{author}{\bibfnamefont{K.}~\bibnamefont{Sacha}},
  \emph{\bibinfo{title}{Time Crystals}} (\bibinfo{publisher}{Springer
  International Publishing}, \bibinfo{address}{Switzerland, Cham},
  \bibinfo{year}{2020}), ISBN \bibinfo{isbn}{978-3-030-52523-1},
  \urlprefix\url{https://doi.org/10.1007/978-3-030-52523-1}.

\bibitem[{\citenamefont{Guo}(2021)}]{GuoBook2021}
\bibinfo{author}{\bibfnamefont{L.}~\bibnamefont{Guo}},
  \emph{\bibinfo{title}{Phase Space Crystals}}, 2053-2563
  (\bibinfo{publisher}{IOP Publishing}, \bibinfo{year}{2021}), ISBN
  \bibinfo{isbn}{978-0-7503-3563-8},
  \urlprefix\url{https://dx.doi.org/10.1088/978-0-7503-3563-8}.

\bibitem[{\citenamefont{Buchleitner et~al.}(2002)\citenamefont{Buchleitner,
  Delande, and Zakrzewski}}]{Buchleitner2002}
\bibinfo{author}{\bibfnamefont{A.}~\bibnamefont{Buchleitner}},
  \bibinfo{author}{\bibfnamefont{D.}~\bibnamefont{Delande}}, \bibnamefont{and}
  \bibinfo{author}{\bibfnamefont{J.}~\bibnamefont{Zakrzewski}},
  \bibinfo{journal}{Physics reports} \textbf{\bibinfo{volume}{368}},
  \bibinfo{pages}{409} (\bibinfo{year}{2002}),
  \urlprefix\url{http://www.sciencedirect.com/science/article/pii/S0370157302002703}.

\bibitem[{\citenamefont{Delande and Buchleitner}(1994)}]{Delande1994}
\bibinfo{author}{\bibfnamefont{D.}~\bibnamefont{Delande}} \bibnamefont{and}
  \bibinfo{author}{\bibfnamefont{A.}~\bibnamefont{Buchleitner}}, in
  \emph{\bibinfo{booktitle}{{}}}, edited by
  \bibinfo{editor}{\bibfnamefont{B.}~\bibnamefont{Bederson}} \bibnamefont{and}
  \bibinfo{editor}{\bibfnamefont{H.}~\bibnamefont{Walther}}
  (\bibinfo{publisher}{Academic Press}, \bibinfo{year}{1994}),
  vol.~\bibinfo{volume}{34} of \emph{\bibinfo{series}{{Advances In Atomic,
  Molecular, and Optical Physics}}}, pp. \bibinfo{pages}{85--123},
  \urlprefix\url{http://www.sciencedirect.com/science/article/pii/S1049250X08600750}.

\bibitem[{\citenamefont{Bialynicki-Birula
  et~al.}(1994)\citenamefont{Bialynicki-Birula, Kali\ifmmode~\acute{n}\else
  \'{n}\fi{}ski, and Eberly}}]{Bialynicki1994}
\bibinfo{author}{\bibfnamefont{I.}~\bibnamefont{Bialynicki-Birula}},
  \bibinfo{author}{\bibfnamefont{M.}~\bibnamefont{Kali\ifmmode~\acute{n}\else
  \'{n}\fi{}ski}}, \bibnamefont{and} \bibinfo{author}{\bibfnamefont{J.~H.}
  \bibnamefont{Eberly}}, \bibinfo{journal}{Phys. Rev. Lett.}
  \textbf{\bibinfo{volume}{73}}, \bibinfo{pages}{1777} (\bibinfo{year}{1994}),
  \urlprefix\url{https://link.aps.org/doi/10.1103/PhysRevLett.73.1777}.

\bibitem[{\citenamefont{Maeda and Gallagher}(2004)}]{Maeda2004}
\bibinfo{author}{\bibfnamefont{H.}~\bibnamefont{Maeda}} \bibnamefont{and}
  \bibinfo{author}{\bibfnamefont{T.~F.} \bibnamefont{Gallagher}},
  \bibinfo{journal}{Phys. Rev. Lett.} \textbf{\bibinfo{volume}{92}},
  \bibinfo{pages}{133004} (\bibinfo{year}{2004}),
  \urlprefix\url{http://link.aps.org/doi/10.1103/PhysRevLett.92.133004}.

\bibitem[{\citenamefont{Maeda and Gallagher}(2007)}]{Maeda2007}
\bibinfo{author}{\bibfnamefont{H.}~\bibnamefont{Maeda}} \bibnamefont{and}
  \bibinfo{author}{\bibfnamefont{T.~F.} \bibnamefont{Gallagher}},
  \bibinfo{journal}{Phys. Rev. A} \textbf{\bibinfo{volume}{75}},
  \bibinfo{pages}{033410} (\bibinfo{year}{2007}),
  \urlprefix\url{https://link.aps.org/doi/10.1103/PhysRevA.75.033410}.

\bibitem[{\citenamefont{Maeda et~al.}(2009)\citenamefont{Maeda, Gurian, and
  Gallagher}}]{Maeda2009}
\bibinfo{author}{\bibfnamefont{H.}~\bibnamefont{Maeda}},
  \bibinfo{author}{\bibfnamefont{J.~H.} \bibnamefont{Gurian}},
  \bibnamefont{and} \bibinfo{author}{\bibfnamefont{T.~F.}
  \bibnamefont{Gallagher}}, \bibinfo{journal}{Phys. Rev. Lett.}
  \textbf{\bibinfo{volume}{102}}, \bibinfo{pages}{103001}
  (\bibinfo{year}{2009}),
  \urlprefix\url{https://link.aps.org/doi/10.1103/PhysRevLett.102.103001}.

\bibitem[{\citenamefont{Wyker et~al.}(2012)\citenamefont{Wyker, Ye, Dunning,
  Yoshida, Reinhold, and Burgd\"orfer}}]{Wyker2012}
\bibinfo{author}{\bibfnamefont{B.}~\bibnamefont{Wyker}},
  \bibinfo{author}{\bibfnamefont{S.}~\bibnamefont{Ye}},
  \bibinfo{author}{\bibfnamefont{F.~B.} \bibnamefont{Dunning}},
  \bibinfo{author}{\bibfnamefont{S.}~\bibnamefont{Yoshida}},
  \bibinfo{author}{\bibfnamefont{C.~O.} \bibnamefont{Reinhold}},
  \bibnamefont{and}
  \bibinfo{author}{\bibfnamefont{J.}~\bibnamefont{Burgd\"orfer}},
  \bibinfo{journal}{Phys. Rev. Lett.} \textbf{\bibinfo{volume}{108}},
  \bibinfo{pages}{043001} (\bibinfo{year}{2012}),
  \urlprefix\url{https://link.aps.org/doi/10.1103/PhysRevLett.108.043001}.

\bibitem[{\citenamefont{Giergiel and Sacha}(2017)}]{Giergiel2017}
\bibinfo{author}{\bibfnamefont{K.}~\bibnamefont{Giergiel}} \bibnamefont{and}
  \bibinfo{author}{\bibfnamefont{K.}~\bibnamefont{Sacha}},
  \bibinfo{journal}{Phys. Rev. A} \textbf{\bibinfo{volume}{95}},
  \bibinfo{pages}{063402} (\bibinfo{year}{2017}),
  \urlprefix\url{https://link.aps.org/doi/10.1103/PhysRevA.95.063402}.

\bibitem[{\citenamefont{{Eiles} et~al.}(2021)\citenamefont{{Eiles}, {Eisfeld},
  and {Rost}}}]{Eiles2021}
\bibinfo{author}{\bibfnamefont{M.~T.} \bibnamefont{{Eiles}}},
  \bibinfo{author}{\bibfnamefont{A.}~\bibnamefont{{Eisfeld}}},
  \bibnamefont{and} \bibinfo{author}{\bibfnamefont{J.~M.}
  \bibnamefont{{Rost}}}, \bibinfo{journal}{arXiv e-prints}
  \bibinfo{eid}{arXiv:2111.10345} (\bibinfo{year}{2021}), \eprint{2111.10345}.

\bibitem[{\citenamefont{Lichtenberg and Lieberman}(1992)}]{Lichtenberg1992}
\bibinfo{author}{\bibfnamefont{A.}~\bibnamefont{Lichtenberg}} \bibnamefont{and}
  \bibinfo{author}{\bibfnamefont{M.}~\bibnamefont{Lieberman}},
  \emph{\bibinfo{title}{Regular and chaotic dynamics}}, Applied mathematical
  sciences (\bibinfo{publisher}{Springer-Verlag}, \bibinfo{year}{1992}), ISBN
  \bibinfo{isbn}{9783540977452},
  \urlprefix\url{https://books.google.pl/books?id=2ssPAQAAMAAJ}.

\bibitem[{SM()}]{SM}
\bibinfo{note}{See Supplemental Material where the details of the deriviation
  of the classical and quantum effective Hamiltonians and the discussion of
  robustness of the realization of the effective models are presented.}

\bibitem[{\citenamefont{{Sacha, K.} et~al.}(1998)\citenamefont{{Sacha, K.},
  {Zakrzewski, J.}, and {Delande, D.}}}]{Sacha1998}
\bibinfo{author}{\bibnamefont{{Sacha, K.}}},
  \bibinfo{author}{\bibnamefont{{Zakrzewski, J.}}}, \bibnamefont{and}
  \bibinfo{author}{\bibnamefont{{Delande, D.}}}, \bibinfo{journal}{Eur. Phys.
  J. D} \textbf{\bibinfo{volume}{1}}, \bibinfo{pages}{231}
  (\bibinfo{year}{1998}),
  \urlprefix\url{https://doi.org/10.1007/s100530050086}.

\bibitem[{\citenamefont{Su et~al.}(1979)\citenamefont{Su, Schrieffer, and
  Heeger}}]{Su1979}
\bibinfo{author}{\bibfnamefont{W.~P.} \bibnamefont{Su}},
  \bibinfo{author}{\bibfnamefont{J.~R.} \bibnamefont{Schrieffer}},
  \bibnamefont{and} \bibinfo{author}{\bibfnamefont{A.~J.}
  \bibnamefont{Heeger}}, \bibinfo{journal}{Phys. Rev. Lett.}
  \textbf{\bibinfo{volume}{42}}, \bibinfo{pages}{1698} (\bibinfo{year}{1979}),
  \urlprefix\url{https://link.aps.org/doi/10.1103/PhysRevLett.42.1698}.

\bibitem[{\citenamefont{Asb{\'o}th et~al.}(2016)\citenamefont{Asb{\'o}th,
  Oroszl{\'a}ny, and P{\'a}lyi}}]{Asboth2016short}
\bibinfo{author}{\bibfnamefont{J.}~\bibnamefont{Asb{\'o}th}},
  \bibinfo{author}{\bibfnamefont{L.}~\bibnamefont{Oroszl{\'a}ny}},
  \bibnamefont{and}
  \bibinfo{author}{\bibfnamefont{A.}~\bibnamefont{P{\'a}lyi}},
  \emph{\bibinfo{title}{A Short Course on Topological Insulators: Band
  Structure and Edge States in One and Two Dimensions}}, Lecture Notes in
  Physics (\bibinfo{publisher}{Springer International Publishing},
  \bibinfo{year}{2016}), ISBN \bibinfo{isbn}{9783319256078},
  \urlprefix\url{https://books.google.pl/books?id=RWKhCwAAQBAJ}.

\bibitem[{\citenamefont{Cooper et~al.}(2019)\citenamefont{Cooper, Dalibard, and
  Spielman}}]{Cooper2019}
\bibinfo{author}{\bibfnamefont{N.~R.} \bibnamefont{Cooper}},
  \bibinfo{author}{\bibfnamefont{J.}~\bibnamefont{Dalibard}}, \bibnamefont{and}
  \bibinfo{author}{\bibfnamefont{I.~B.} \bibnamefont{Spielman}},
  \bibinfo{journal}{Rev. Mod. Phys.} \textbf{\bibinfo{volume}{91}},
  \bibinfo{pages}{015005} (\bibinfo{year}{2019}),
  \urlprefix\url{https://link.aps.org/doi/10.1103/RevModPhys.91.015005}.

\bibitem[{\citenamefont{Atala et~al.}(2013)\citenamefont{Atala, Aidelsburger,
  Barreiro, Abanin, Kitagawa, Demler, and Bloch}}]{Atala2013}
\bibinfo{author}{\bibfnamefont{M.}~\bibnamefont{Atala}},
  \bibinfo{author}{\bibfnamefont{M.}~\bibnamefont{Aidelsburger}},
  \bibinfo{author}{\bibfnamefont{J.~T.} \bibnamefont{Barreiro}},
  \bibinfo{author}{\bibfnamefont{D.}~\bibnamefont{Abanin}},
  \bibinfo{author}{\bibfnamefont{T.}~\bibnamefont{Kitagawa}},
  \bibinfo{author}{\bibfnamefont{E.}~\bibnamefont{Demler}}, \bibnamefont{and}
  \bibinfo{author}{\bibfnamefont{I.}~\bibnamefont{Bloch}},
  \bibinfo{journal}{Nature Physics} \textbf{\bibinfo{volume}{9}},
  \bibinfo{pages}{795} (\bibinfo{year}{2013}), ISSN \bibinfo{issn}{1745-2481},
  \urlprefix\url{https://doi.org/10.1038/nphys2790}.

\bibitem[{\citenamefont{{Meier Eric J.} et~al.}(2016)\citenamefont{{Meier Eric
  J.}, {An Fangzhao Alex}, and {Gadway Bryce}}}]{Meier2016}
\bibinfo{author}{\bibnamefont{{Meier Eric J.}}},
  \bibinfo{author}{\bibnamefont{{An Fangzhao Alex}}}, \bibnamefont{and}
  \bibinfo{author}{\bibnamefont{{Gadway Bryce}}}, \bibinfo{journal}{Nature
  Communications} \textbf{\bibinfo{volume}{7}}, \bibinfo{pages}{13986}
  (\bibinfo{year}{2016}),
  \urlprefix\url{https://www.nature.com/articles/ncomms13986\#supplementary-information}.

\bibitem[{\citenamefont{St-Jean et~al.}(2017)\citenamefont{St-Jean, Goblot,
  Galopin, Lema{\^i}tre, Ozawa, Le~Gratiet, Sagnes, Bloch, and
  Amo}}]{St-Jean2017}
\bibinfo{author}{\bibfnamefont{P.}~\bibnamefont{St-Jean}},
  \bibinfo{author}{\bibfnamefont{V.}~\bibnamefont{Goblot}},
  \bibinfo{author}{\bibfnamefont{E.}~\bibnamefont{Galopin}},
  \bibinfo{author}{\bibfnamefont{A.}~\bibnamefont{Lema{\^i}tre}},
  \bibinfo{author}{\bibfnamefont{T.}~\bibnamefont{Ozawa}},
  \bibinfo{author}{\bibfnamefont{L.}~\bibnamefont{Le~Gratiet}},
  \bibinfo{author}{\bibfnamefont{I.}~\bibnamefont{Sagnes}},
  \bibinfo{author}{\bibfnamefont{J.}~\bibnamefont{Bloch}}, \bibnamefont{and}
  \bibinfo{author}{\bibfnamefont{A.}~\bibnamefont{Amo}},
  \bibinfo{journal}{Nature Photonics} \textbf{\bibinfo{volume}{11}},
  \bibinfo{pages}{651} (\bibinfo{year}{2017}), ISSN \bibinfo{issn}{1749-4893},
  \urlprefix\url{https://doi.org/10.1038/s41566-017-0006-2}.

\bibitem[{\citenamefont{Giergiel et~al.}(2019)\citenamefont{Giergiel, Dauphin,
  Lewenstein, Zakrzewski, and Sacha}}]{Giergiel2018b}
\bibinfo{author}{\bibfnamefont{K.}~\bibnamefont{Giergiel}},
  \bibinfo{author}{\bibfnamefont{A.}~\bibnamefont{Dauphin}},
  \bibinfo{author}{\bibfnamefont{M.}~\bibnamefont{Lewenstein}},
  \bibinfo{author}{\bibfnamefont{J.}~\bibnamefont{Zakrzewski}},
  \bibnamefont{and} \bibinfo{author}{\bibfnamefont{K.}~\bibnamefont{Sacha}},
  \bibinfo{journal}{New Journal of Physics} \textbf{\bibinfo{volume}{21}},
  \bibinfo{pages}{052003} (\bibinfo{year}{2019}),
  \urlprefix\url{https://doi.org/10.1088/1367-2630/ab1e5f}.

\bibitem[{foo()}]{foot1}
\bibinfo{note}{The chosen Fourier components $f_k$ allow one to realize a
  barrier in the effective model (2). This barrier can be realized in many
  different ways, see \cite{SM}. The small parameter $\epsilon$ breaks the
  reflection symmetry of the potential in the effective SSH model --- if such a
  symmetry was present, the edge states would be superpositions of the
  localized wavepackets presented in Fig.~1(b).}

\bibitem[{\citenamefont{Giergiel et~al.}(2018)\citenamefont{Giergiel,
  Miroszewski, and Sacha}}]{Giergiel2018}
\bibinfo{author}{\bibfnamefont{K.}~\bibnamefont{Giergiel}},
  \bibinfo{author}{\bibfnamefont{A.}~\bibnamefont{Miroszewski}},
  \bibnamefont{and} \bibinfo{author}{\bibfnamefont{K.}~\bibnamefont{Sacha}},
  \bibinfo{journal}{Phys. Rev. Lett.} \textbf{\bibinfo{volume}{120}},
  \bibinfo{pages}{140401} (\bibinfo{year}{2018}),
  \urlprefix\url{https://link.aps.org/doi/10.1103/PhysRevLett.120.140401}.

\bibitem[{\citenamefont{Matus et~al.}(2021)\citenamefont{Matus, Giergiel, and
  Sacha}}]{Matus2021}
\bibinfo{author}{\bibfnamefont{P.}~\bibnamefont{Matus}},
  \bibinfo{author}{\bibfnamefont{K.}~\bibnamefont{Giergiel}}, \bibnamefont{and}
  \bibinfo{author}{\bibfnamefont{K.}~\bibnamefont{Sacha}},
  \bibinfo{journal}{Phys. Rev. A} \textbf{\bibinfo{volume}{103}},
  \bibinfo{pages}{023320} (\bibinfo{year}{2021}),
  \urlprefix\url{https://link.aps.org/doi/10.1103/PhysRevA.103.023320}.

\bibitem[{\citenamefont{Gaunt et~al.}(2013)\citenamefont{Gaunt, Schmidutz,
  Gotlibovych, Smith, and Hadzibabic}}]{Gaunt2013}
\bibinfo{author}{\bibfnamefont{A.~L.} \bibnamefont{Gaunt}},
  \bibinfo{author}{\bibfnamefont{T.~F.} \bibnamefont{Schmidutz}},
  \bibinfo{author}{\bibfnamefont{I.}~\bibnamefont{Gotlibovych}},
  \bibinfo{author}{\bibfnamefont{R.~P.} \bibnamefont{Smith}}, \bibnamefont{and}
  \bibinfo{author}{\bibfnamefont{Z.}~\bibnamefont{Hadzibabic}},
  \bibinfo{journal}{Phys. Rev. Lett.} \textbf{\bibinfo{volume}{110}},
  \bibinfo{pages}{200406} (\bibinfo{year}{2013}),
  \urlprefix\url{https://link.aps.org/doi/10.1103/PhysRevLett.110.200406}.

\bibitem[{\citenamefont{Chin et~al.}(2010)\citenamefont{Chin, Grimm, Julienne,
  and Tiesinga}}]{Chin2010}
\bibinfo{author}{\bibfnamefont{C.}~\bibnamefont{Chin}},
  \bibinfo{author}{\bibfnamefont{R.}~\bibnamefont{Grimm}},
  \bibinfo{author}{\bibfnamefont{P.}~\bibnamefont{Julienne}}, \bibnamefont{and}
  \bibinfo{author}{\bibfnamefont{E.}~\bibnamefont{Tiesinga}},
  \bibinfo{journal}{Rev. Mod. Phys.} \textbf{\bibinfo{volume}{82}},
  \bibinfo{pages}{1225} (\bibinfo{year}{2010}),
  \urlprefix\url{https://link.aps.org/doi/10.1103/RevModPhys.82.1225}.

\bibitem[{\citenamefont{Olshanii}(1998)}]{Olshanii1998}
\bibinfo{author}{\bibfnamefont{M.}~\bibnamefont{Olshanii}},
  \bibinfo{journal}{Phys. Rev. Lett.} \textbf{\bibinfo{volume}{81}},
  \bibinfo{pages}{938} (\bibinfo{year}{1998}),
  \urlprefix\url{https://link.aps.org/doi/10.1103/PhysRevLett.81.938}.

\bibitem[{\citenamefont{Shirley}(1965)}]{Shirley1965}
\bibinfo{author}{\bibfnamefont{J.~H.} \bibnamefont{Shirley}},
  \bibinfo{journal}{Phys. Rev.} \textbf{\bibinfo{volume}{138}},
  \bibinfo{pages}{B979} (\bibinfo{year}{1965}),
  \urlprefix\url{https://link.aps.org/doi/10.1103/PhysRev.138.B979}.

\bibitem[{\citenamefont{Haller et~al.}(2010)\citenamefont{Haller, Mark, Hart,
  Danzl, Reichs\"ollner, Melezhik, Schmelcher, and N\"agerl}}]{Haller2010}
\bibinfo{author}{\bibfnamefont{E.}~\bibnamefont{Haller}},
  \bibinfo{author}{\bibfnamefont{M.~J.} \bibnamefont{Mark}},
  \bibinfo{author}{\bibfnamefont{R.}~\bibnamefont{Hart}},
  \bibinfo{author}{\bibfnamefont{J.~G.} \bibnamefont{Danzl}},
  \bibinfo{author}{\bibfnamefont{L.}~\bibnamefont{Reichs\"ollner}},
  \bibinfo{author}{\bibfnamefont{V.}~\bibnamefont{Melezhik}},
  \bibinfo{author}{\bibfnamefont{P.}~\bibnamefont{Schmelcher}},
  \bibnamefont{and} \bibinfo{author}{\bibfnamefont{H.-C.}
  \bibnamefont{N\"agerl}}, \bibinfo{journal}{Phys. Rev. Lett.}
  \textbf{\bibinfo{volume}{104}}, \bibinfo{pages}{153203}
  (\bibinfo{year}{2010}),
  \urlprefix\url{https://link.aps.org/doi/10.1103/PhysRevLett.104.153203}.

\bibitem[{\citenamefont{Balay et~al.}(2019)\citenamefont{Balay, Abhyankar,
  Adams, Brown, Brune, Buschelman, Dalcin, Dener, Eijkhout, Gropp
  et~al.}}]{petsc-user-ref}
\bibinfo{author}{\bibfnamefont{S.}~\bibnamefont{Balay}},
  \bibinfo{author}{\bibfnamefont{S.}~\bibnamefont{Abhyankar}},
  \bibinfo{author}{\bibfnamefont{M.~F.} \bibnamefont{Adams}},
  \bibinfo{author}{\bibfnamefont{J.}~\bibnamefont{Brown}},
  \bibinfo{author}{\bibfnamefont{P.}~\bibnamefont{Brune}},
  \bibinfo{author}{\bibfnamefont{K.}~\bibnamefont{Buschelman}},
  \bibinfo{author}{\bibfnamefont{L.}~\bibnamefont{Dalcin}},
  \bibinfo{author}{\bibfnamefont{A.}~\bibnamefont{Dener}},
  \bibinfo{author}{\bibfnamefont{V.}~\bibnamefont{Eijkhout}},
  \bibinfo{author}{\bibfnamefont{W.~D.} \bibnamefont{Gropp}},
  \bibnamefont{et~al.}, \bibinfo{type}{Tech. Rep.} \bibinfo{number}{ANL-95/11 -
  Revision 3.12}, \bibinfo{institution}{Argonne National Laboratory}
  (\bibinfo{year}{2019}), \urlprefix\url{https://www.mcs.anl.gov/petsc}.

\bibitem[{\citenamefont{Balay et~al.}(1997)\citenamefont{Balay, Gropp, McInnes,
  and Smith}}]{petsc-efficient}
\bibinfo{author}{\bibfnamefont{S.}~\bibnamefont{Balay}},
  \bibinfo{author}{\bibfnamefont{W.~D.} \bibnamefont{Gropp}},
  \bibinfo{author}{\bibfnamefont{L.~C.} \bibnamefont{McInnes}},
  \bibnamefont{and} \bibinfo{author}{\bibfnamefont{B.~F.} \bibnamefont{Smith}},
  in \emph{\bibinfo{booktitle}{Modern Software Tools in Scientific Computing}},
  edited by \bibinfo{editor}{\bibfnamefont{E.}~\bibnamefont{Arge}},
  \bibinfo{editor}{\bibfnamefont{A.~M.} \bibnamefont{Bruaset}},
  \bibnamefont{and} \bibinfo{editor}{\bibfnamefont{H.~P.}
  \bibnamefont{Langtangen}} (\bibinfo{publisher}{Birkh{\"{a}}user Press},
  \bibinfo{year}{1997}), pp. \bibinfo{pages}{163--202}.

\bibitem[{\citenamefont{Hernandez et~al.}(2005)\citenamefont{Hernandez, Roman,
  and Vidal}}]{Hernandez:2005}
\bibinfo{author}{\bibfnamefont{V.}~\bibnamefont{Hernandez}},
  \bibinfo{author}{\bibfnamefont{J.~E.} \bibnamefont{Roman}}, \bibnamefont{and}
  \bibinfo{author}{\bibfnamefont{V.}~\bibnamefont{Vidal}},
  \bibinfo{journal}{ACM Trans. Math. Softw.} \textbf{\bibinfo{volume}{31}},
  \bibinfo{pages}{351} (\bibinfo{year}{2005}), ISSN \bibinfo{issn}{0098-3500},
  \urlprefix\url{http://doi.acm.org/10.1145/1089014.1089019}.

\bibitem[{\citenamefont{Guo and Marthaler}(2016)}]{Guo2016}
\bibinfo{author}{\bibfnamefont{L.}~\bibnamefont{Guo}} \bibnamefont{and}
  \bibinfo{author}{\bibfnamefont{M.}~\bibnamefont{Marthaler}},
  \bibinfo{journal}{New Journal of Physics} \textbf{\bibinfo{volume}{18}},
  \bibinfo{pages}{023006} (\bibinfo{year}{2016}),
  \urlprefix\url{http://stacks.iop.org/1367-2630/18/i=2/a=023006}.

\bibitem[{\citenamefont{Tanzi et~al.}(2018)\citenamefont{Tanzi, Cabrera, Sanz,
  Cheiney, Tomza, and Tarruell}}]{Tanzi2018}
\bibinfo{author}{\bibfnamefont{L.}~\bibnamefont{Tanzi}},
  \bibinfo{author}{\bibfnamefont{C.~R.} \bibnamefont{Cabrera}},
  \bibinfo{author}{\bibfnamefont{J.}~\bibnamefont{Sanz}},
  \bibinfo{author}{\bibfnamefont{P.}~\bibnamefont{Cheiney}},
  \bibinfo{author}{\bibfnamefont{M.}~\bibnamefont{Tomza}}, \bibnamefont{and}
  \bibinfo{author}{\bibfnamefont{L.}~\bibnamefont{Tarruell}},
  \bibinfo{journal}{Phys. Rev. A} \textbf{\bibinfo{volume}{98}},
  \bibinfo{pages}{062712} (\bibinfo{year}{2018}),
  \urlprefix\url{https://link.aps.org/doi/10.1103/PhysRevA.98.062712}.

\end{thebibliography}

\section{\large Supplemental Materials}

\setcounter{equation}{0}
\setcounter{figure}{0}
\setcounter{table}{0}
\makeatletter
\renewcommand{\theequation}{S\arabic{equation}}
\renewcommand{\thefigure}{S\arabic{figure}}

In this Supplemental Material, we derive the classical and quantum effective Hamiltonians and discuss robustness of the realization of the effective models. 

\section{Hydrogen atom}
\label{SHydrogen}

We consider a H atom in the presence of static electric and linearly polarized microwave fields which in the atomic units is descrbed by the Hamiltonian
\be
H=\frac{\vect{p}^2}{2}-\frac{1}{r}+z[F+F_1\cos(\omega t)+F_2\cos(2\omega t)+\lambda f(t)],
\label{Sh}
\ee
where
\be
f(t)=f(t+s2\pi/\omega)=\sum_{\substack{k\ne 0}}f_ke^{ik\omega t/s},
\label{Sf}
\ee
$F$, $F_{j}$ and $\lambda$ are the strength of the electric field and the amplitudes of the microwaves, respectively, and $s$ is an integer number.

\begin{figure}[t]
\includegraphics[width=.95\columnwidth]{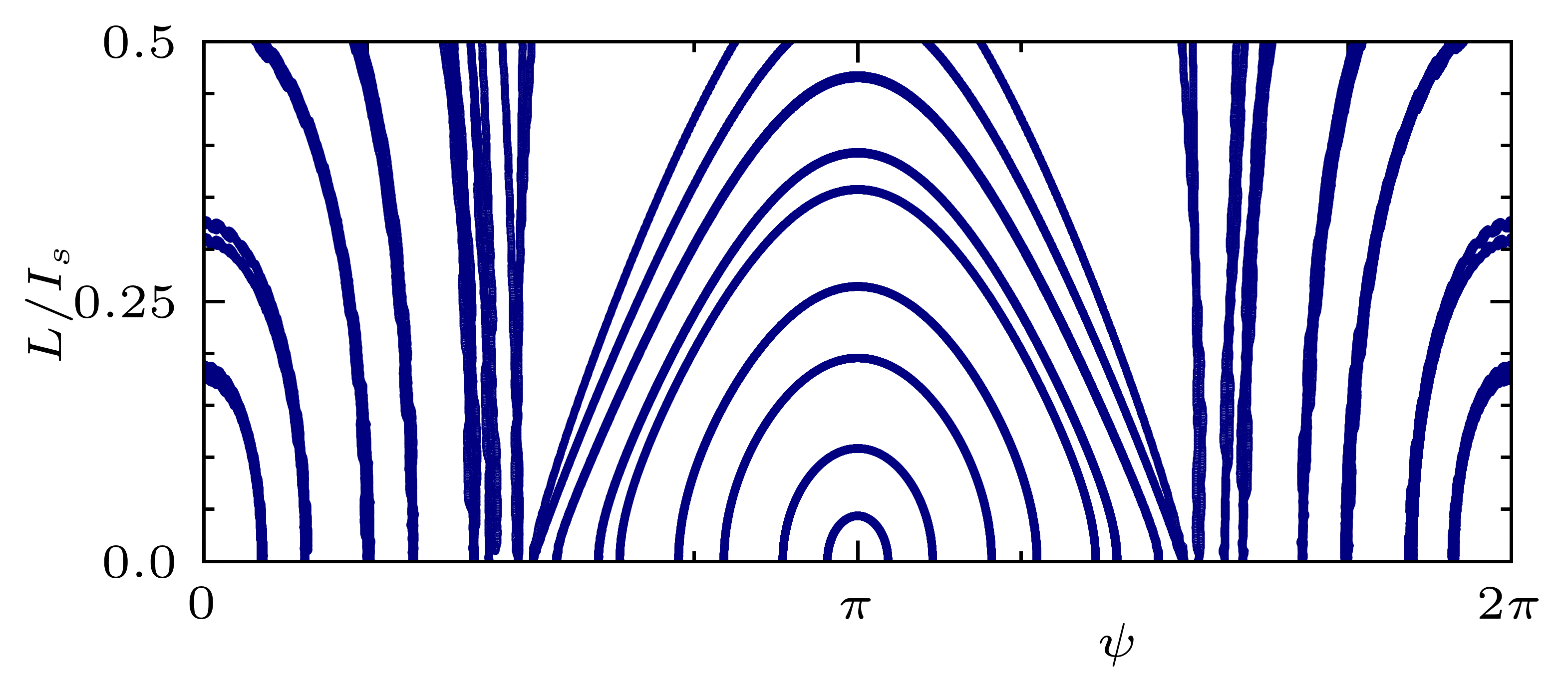}
\caption{Hydrogen atom. Stroboscopic map in the $(L,\psi)$ space obtained by integrating the exact equations of motion for different initial values $L$ and $\psi$ and for the initial values of $I/I_s=1$ and $\theta=0$. The parameters of the system are the following: $s=I_s^3\omega=4$, $I_s^4F=1.5\times10^{-4}$, $I_s^4F_1=1.258\times10^{-3}$, $I_s^4F_2=1.93\times10^{-3}$, and $\lambda=1.172\times10^{-5}$. The stable fixed point is located at $L=0$ and $\psi=0$ if the static electric field strength is $I_s^4F\gtrapprox 4\times10^{-5}$. This fixed point corresponds to the resonant Kepler ellipse degenerated into a line that is oriented along the $z$ axis.}
\label{SLpsi}
\end{figure}

\begin{figure}[t]
\includegraphics[width=.95\columnwidth]{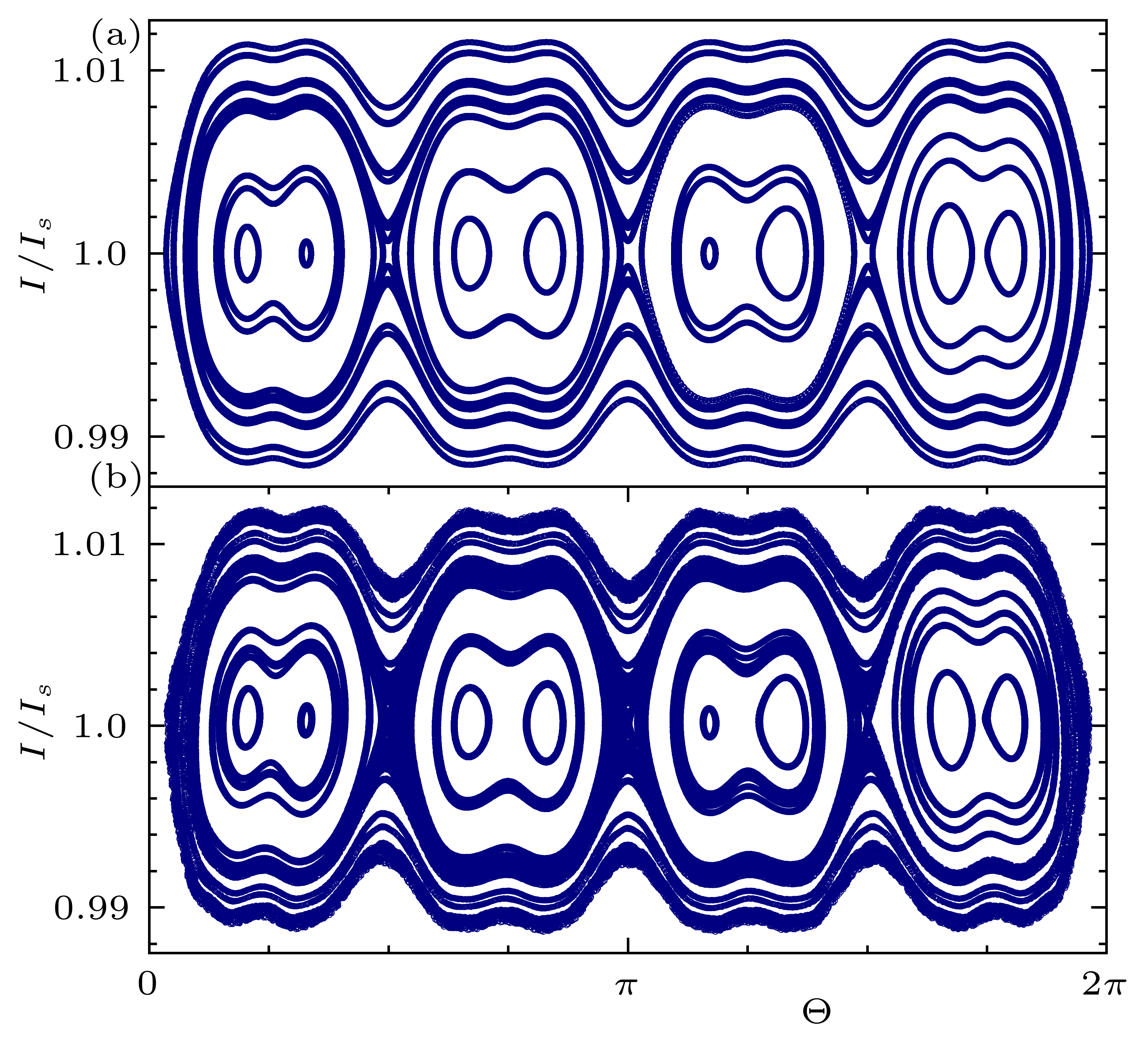}
\caption{Hydrogen atom. Panel~(a): phase space portrait $(I,\Theta)$ obtained by means of (\ref{Sheff}). Panel~(b): the corresponding exact stroboscopic map of 3D hydrogen obtained for the initial values of $L/I_s=0.1$, $\psi=0$ and the different initial values of $I$ and $\Theta$. The parameters of the system are the same as in Fig.~\ref{SLpsi}.}
\label{Sfig1}
\end{figure}

Let us begin with the classical description. In terms of the so-called action-angle variables \cite{Lichtenberg1992,Buchleitner2002}, the Hamiltonian (\ref{Sh}) reads
\bea
H&=& -\frac{1}{2I^2}+[F+F_1\cos(\omega t)+F_2\cos(2\omega t)+\lambda f(t)]
\cr && \times \sum_{m=-\infty}^{+\infty} U_m(I,L) \cos(m\theta +\psi),
\eea
where
\bea
U_m(I,L)&=&-\frac{I^2}{m}\left[{\cal J}_m'(m\tilde e)+\frac{\sqrt{1-\tilde e^2}}{\tilde e}{\cal J}_m(m\tilde e)\right], \quad m\ne 0,
\cr
U_0(I)&=&\frac32 \tilde e I^2,
\cr
\tilde e&=& \sqrt{1-\frac{L^2}{I^2}}.
\eea
$\tilde e$ is the eccentricity of a Kepler orbit of the Rydberg electron, and ${\cal J}_m$, ${\cal J}_m'$ denote the Bessel functions and their derivatives, respectively.
We have assumed that the projection of the angular momentum of the electron on the $z$ axis (which is a constant of motion) is zero. The principal action $I$ is the classical analogue of the principal quantum number $n$ and the conjugate angle $\theta$ describes the position of the electron in a Kepler orbit. The angular momentum $L$ is conjugate to $\psi$ which is the angle between the major axis of a Kepler ellipse and the $z$ axis. We are going to describe the electron close to the resonant orbit where the microwave frequency $\omega$ is $s$ times higher than the frequency of the unperturbed electronic motion given by $1/I^3$ [$s$ is the same as in (\ref{Sf})]. If we switch to the moving frame,
\be
\Theta=\theta-\frac{\omega}{s}t,
\label{movframe}
\ee
then for $I\approx I_s$, where $\omega I_s^3=s$, all dynamical variables are slowly varying in the limit of the weak perturbation and we may average the Hamiltonian over time keeping $I$, $\Theta$, $L$ and $\psi$ fixed which results in the secular effective Hamiltonian \cite{Lichtenberg1992}
\bea
H_{\rm eff}&\approx& -\frac{1}{2I^2}-\frac{\omega I}{s}+FU_0\cos\psi
\cr
&& + \frac{F_1}{4} \left(U_se^{i\psi}+U_{-s}e^{-i\psi}\right)e^{is\Theta}
\cr
&&+\frac{F_1}{4} \left(U_se^{-i\psi}+U_{-s}e^{i\psi}\right)e^{-is\Theta}
\cr
&& + \frac{F_2}{4} \left(U_{2s}e^{i\psi}+U_{-2s}e^{-i\psi}\right)e^{i2s\Theta}
\cr
&&+\frac{F_2}{4} \left(U_{2s}e^{-i\psi}+U_{-2s}e^{i\psi}\right)e^{-i2s\Theta}
\cr
&& +\frac{\lambda}{2}\sum_{m=-\infty}^{+\infty}\left(  U_{m}e^{i\psi}%
f_{-m}+U_{-m}e^{-i\psi}f_{m}\right)  e^{im\Theta}.
\cr &&
\label{Shav}
\eea
The static electric field controls the shape of stable resonant orbits \cite{Sacha1998}. If the static field dominates over the microwave fields, the stable orbit is the Kepler ellipse degenerate into a line where the resonantly driven electron is moving with the period $s2\pi/\omega$ along the direction of the electric field vector with $L=0$ and $\psi=0$, see Fig.~\ref{SLpsi}. Then electronic motion close to such a resonant orbit can be described by a simplified version of (\ref{Shav})
\be
H_{\rm eff}=\frac{P^2}{2m_{\rm eff}}+V_1\cos(s\Theta)+V_2\cos(2s\Theta)+\lambda V_{\rm b}(\Theta),
\label{Sheff}
\ee
where we set $L=0$ and $\psi=0$ and expanded the first two terms of (\ref{Shav}) in the Taylor series up to the second order in $P=I-I_s$, which is small if we choose initial conditions of the electron close to the resonant orbit. The effective mass $m_{\rm eff}=-I_s^4/3$, $V_{1}=-I_{s}^{2}F_{1}\mathcal{J}_{s}^{\prime}(s)/s$, $V_{2}
=-I_{s}^{2}F_{2}\mathcal{J}_{2s}^{\prime}(2s)/2s$ and $V_{\rm b}(\Theta)=-I_{s}^{2}\sum_{k\neq0}f_{-k}%
\mathcal{J}_{k}^{\prime}(k)e^{ik\Theta}/k$. The effective Hamiltonian (\ref{Sheff}) is the basis of the description of the topological behavior we analyze in the Letter, and its validity can be tested with the help of the exact numerical integration of the classical equations of motion. In Fig.~\ref{Sfig1} we show the phase space portrait generated by (\ref{Sheff}) and the corresponding stroboscopic map obtained by numerically solving the equations of motion generated by the exact Hamiltonian (\ref{Sh}) for $L\approx0$ and $\psi\approx 0$. The agreement is perfect, therefore we may be sure that the exact dynamics reveals the topological behavior we are after.

Quantum description of the electron moving resonantly on the Kepler orbit aligned along the electric field direction can be performed either by quantizing the classical effective Hamiltonian (\ref{Sheff}) or by deriving an effective Hamiltonian fully quantum mechanically. The latter is done by switching to the moving frame with the help of the unitary transformation $e^{i\hat n \omega t/s}$ [which is the quantum analogue of the classical canonical transformation to the moving frame (\ref{movframe})], and averaging the resulting Hamiltonian over time which leads to

\bea
\la n',l'|\hat H_{\rm eff}|n,l\ra &=&\left(-\frac{1}{2n^2}-n\frac{\omega}{s}\right)\delta_{nn'}\delta_{ll'}+\la n',l'|z|n,l\ra
\cr && \times \left[F\delta_{nn'}+F_1(\delta_{n+s,n'}+\delta_{n-s,n'})/2\right.
\cr &&
\left. +F_2(\delta_{n+2s,n'}+\delta_{n-2s,n'})/2
+\lambda f_{n-n'}\right],
\cr &&
\label{Sqheff}
\eea
 where $|n,l\ra$ is the hydrogenic eigenstate with the principal quantum number $n$ ($\hat n|n,l\ra=n|n,l\ra$), the total angular momentum $l$ and the projection of the angular momentum on the $z$ axis equaled zero.
Comparison of the resonant spectra obtained by diagonalization of (\ref{Sqheff}) and (\ref{Sheff}) is presented in Fig.~1 in the Letter. The results match each other very well, which, together with the agreement between the exact and secular classical descriptions (cf. Fig.~\ref{Sfig1}), proves the consistency of the approach we use.

At the end of this section let us discuss robustness of the topologically protected localization of a Rydberg electron. 

The role of the driving function $f(t)$ is to create a localized barrier $V_b(\Theta)$ in the effective potential in Eq.~(\ref{Sheff}). Without $V_b(\Theta)$, there is no edge in the system and no topologically protected localized edge states can be observed. There are basically two requirements that $f(t)$ must meet: (i) the resulting barrier $V_b(\Theta)$ must be localized on a length scale much shorter than the size of the entire crystalline structure and (ii) the barrier height must be larger than the width of the energy bands of the effective model (\ref{Sheff}). The latter condition ensures that the presence of the barrier results in open boundary conditions in our crystalline structure, which implies that there is an edge in the system. 

\begin{figure}[t]
\includegraphics[width=.95\columnwidth]{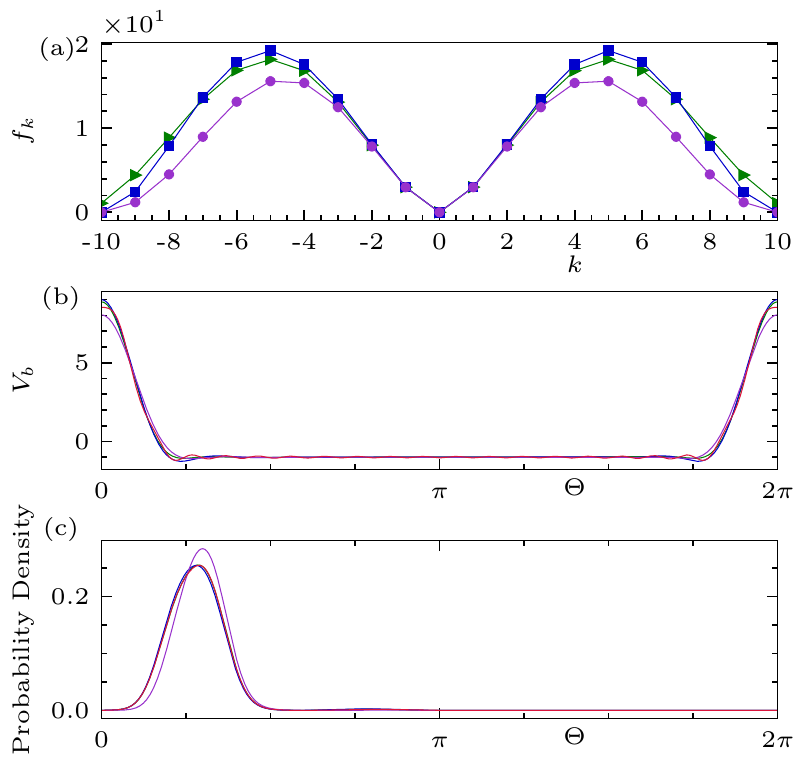}
\caption{Hydrogen atom. Panel~(a) shows different choices of the Fourier components $f_k$ in Eq.~(\ref{Sf}): blue squares correspond to $f_k=\cos^2(k\pi/20)k/J_{k}^{\prime}(k)$, violet circles to $f_k=\sinc^{2}(k\pi/10)k/J_{k}^{\prime}(k)$, and green triangles to $f_k=\cos(k\pi/21)\sinc^{2}(k\pi/14)k/J_{k}^{\prime}(k)$ --- in all cases $f_k=0$ for $|k|>10$. In the Letter $f_k$ presented with green triangles have been chosen but $f_k$ is non-zero for $|k|\le 20$. Lines connecting points are guide for the eye. In panel~(b) plots of the corresponding barrier potentials $V_b(\Theta)$ are shown while in panel~(c) the resulting probability densities of one of the edge states are presented --- the same color coding of the lines as in (a) is chosen. In (b) and (c) the additional red line corresponds to $f_k$ chosen in the Letter. To break the reflection symmetry and obtain the edge states which are not superposition of two wavepackets localized close to $\Theta=0$ and $\Theta=2\pi$, all Fourier components $f_k$ have been multiplied by $e^{ik\epsilon}$ with $\epsilon=5\times 10^{-3}$.}
\label{Sf_of_t}
\end{figure}

The topological localization of a Rydberg electron is stable with respect to different choices of $f(t)$ provided the conditions described above are satisfied. From the experimental point of view, it is important to use $f(t)=\sum_{k}f_ke^{ik\omega t}$ with the smallest possible number of Fourier components $f_k$. The function $f(t)$ used in Fig.~1 in the Letter consists of $f_k$ with $k$ up to $|k_{max}|=20$, but even if we choose $|k_{max}|=10$, the shape of the edge states does not change noticeably, see Fig.~\ref{Sf_of_t}. In addition, the precise functional form of $f(t)$ is not important. In Fig.~\ref{Sf_of_t} we present different choices of $f_k$, the resulting $V_b(\Theta)$ and the probability densities of one of the edge states.

In the experiment, to realize the effective model (\ref{Sheff}) one has to apply a time-periodic driving with frequencies $\omega$ and $2\omega$. This is a critical requirement which, however, can be easily fulfilled in the experiment because having the driving with the frequency $\omega$ its second harmonic with the frequency $2\omega$ can be generated. 

The frequency $\omega$ satisfies the resonant condition $\omega\approx s/n_s^3$ but does not necessarily have to be exactly equal to $s/n_s^3$. If $\omega$ is different from $s/n_s^3$, then the tunneling amplitudes in the resulting Su-Shrieffer-Heeger (SSH \cite{Su1979,Asboth2016short}) model acquire complex phases \cite{Guo2016}. However, these phases can be eliminated by proper redefinition of the phases of the basis states if the barrier potential is present and the open boundary conditions are realized.

It is not necessary to realize the effective model (\ref{Sheff}) with exactly the same parameters as presented in the Letter. The localization of a Rydberg electron is robust, provided that one realizes the effective model in the topologically non-trivial regime. The localization length of the edge states can change with a change in the parameters of the system, but the localization phenomenon itself is robust if the gap between the bands is not closed.

\section{Topological molecules}

In the Letter we introduce the Hamiltonian
\be
H=\frac{p_1^2+p_2^2}{2}+g(t)\delta(x_1-x_2),
\label{Shm}
\ee
which describes motion of two different atoms (the same atomic species but in different hyperfine states) in the squared potential well whose size is $\pi$. We use 
\be
E_0=\frac{\pi^2\hbar^2}{mR^2}, \quad l_0=\frac{R}{\pi},
\label{Sunits}
\ee 
as units of energy and length, respectively, where $R$ is the size of the potential well and $m$ is the mass of the atoms. The strength of contact interactions between atoms is periodically modulated in time
\be
g(t)=F_1\cos(\omega t)+F_2\cos(2\omega t)+\lambda f(t),
\label{Sg}
\ee
where $f(t)$ is given in (\ref{Sf}). We are interested in the motion of the atoms that is resonant with the time-periodic modulation of the interaction strength $g(t)$.

A suitable basis for the quantum description of the atoms is formed by the eigenstates of the non-interacting atoms,
\be
\phi_{n_1,n_2}(x_1,x_2)=\frac{2}{\pi}\sin(n_1 x_1)\sin(n_2 x_2).
\label{Sbasis}
\ee
For antisymmetric combinations of these states, we obtain
\be
\delta(x_1-x_2)\left[\phi_{n_1,n_2}(x_1,x_2)-\phi_{n_2,n_1}(x_1,x_2)\right]=0.
\label{asym_subspace}
\ee
It means that the interaction between the atoms can only take place if they are prepared in the symmetric combinations of (\ref{Sbasis}), i.e.
\bea
\Phi_{n_1,n_2}&=& \frac{1}{\sqrt{2}}\left[\phi_{n_1,n_2}+\phi_{n_2,n_1}\right], \quad n_1>n_2,
\cr
\Phi_{n_1,n_2}&=& \phi_{n_1,n_2}, \quad n_1=n_2,
\label{sym_subspace}
\eea
and thus we may restrict to such a subspace only.

\begin{figure}[t]
\includegraphics[width=.95\columnwidth]{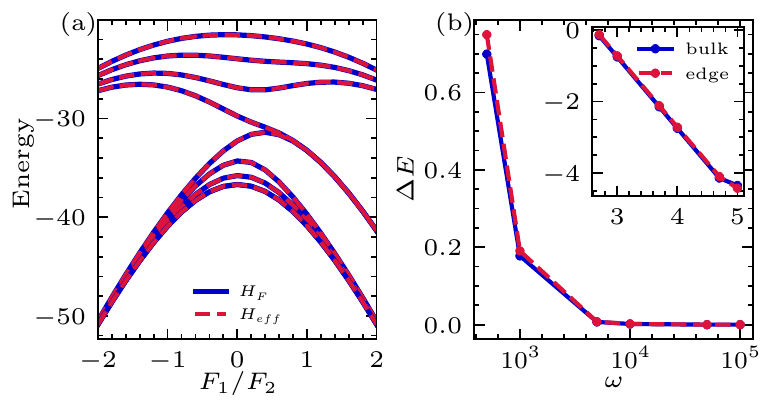}
\caption{Topological molecules. (a) Comparison of the energy levels of the effective Hamiltonian (\ref{Smqheff}) with the corresponding eigenvalues of the exact Floquet Hamiltonian (\ref{SHFnew}) --- only the relevant eigenvalues of $H_F$ are presented. (b) Differences, $\Delta E$, between eigenenergies of the effective Hamiltonian of bulk and edge eigentates, and the corresponding eigenvalues of the Floquet Hamiltonian as a function of $\omega$ in a semi-log plot. Inset shows the same but in a log-log plot. If $\omega$ is sufficiently large, the agreement between the effective and exact results is perfect.
The parameters of the system are the following: $F_2=10$, $\lambda=35$ and Fourier components of $f(t)$ are $f_k=e^{ik \epsilon}\cos(k\pi/21)\sinc^{2}(k\pi/14)$ with $\epsilon=10^{-4}$ for $|k|\le 40$ and $f_k=0$ for greater $|k|$. In (a), $\omega=10^5$ and in (b), $F_1/F_2=2$.}
\label{Sfig3}
\end{figure}

The eigenstates of the effective Hamiltonian (\ref{Smqheff}) are a secular approximation of the exact Floquet states. The resonant motion of the atoms means that $n_1\approx \omega/2s$ and $n_2\approx \omega/2s$ where $s$ is an integer and the same as in (\ref{Sf}). In order to derive an effective quantum Hamiltonian that can describe the resonant behavior of the atoms we switch to the moving frame by means of the unitary transformation $e^{i(\hat n_1+\hat n_2)\omega t/2s}$ and average the resulting Hamiltonian over time. Introducing the quantum numbers $n_{\rm cm}=n_1-n_2$ and $n=(n_1+n_2)/2-\omega/2s$, the matrix elements of the obtained effective Hamiltonian read
\bea
\la n_{\rm cm}',n'|\hat H_{\rm eff}|n_{\rm cm},n\ra
&=&\left[n^2\delta_{n',n}+\frac{F_1}{2\pi}(\delta_{n',n+s}+\delta_{n',n-s})\right.
\cr &&
+\frac{F_2}{2\pi}(\delta_{n',n+2s}+\delta_{n',n-2s})
\cr &&
\left.+\frac{\lambda}{\pi}f_{n'-n}+\frac{n_{\rm cm}^2}{4}\delta_{n',n}\right]\delta_{n_{\rm cm}',n_{\rm cm}},
\cr &&
\label{Smqheff}
\eea
This effective Hamiltonian, which is valid for $|n|,|n'|\ll \omega/2s$, describes a two-particle system whose center of mass degree of freedom behaves like a free particle, i.e. there is only kinetic energy given by $n_{\rm cm}^2/4$.  The relative position degree of freedom of the particles (corresponding to the momentum quantum numbers $n$) behaves in the same way as the Rydberg electron described in the previous section. Indeed, the matrix elements (\ref{Smqheff}) are identical to the matrix elements of the quantized version of the classical effective Hamiltonian (\ref{Sheff}) calculated in the plane wave basis, $e^{in\Theta}/\sqrt{2\pi}$, if $m_{\rm eff}=1/2$, $V_j=F_j/\pi$ and $V_b=\sum_{k\ne0}f_{k}e^{ik\Theta}/\pi$. Thus, the two-particle system can be prepared in a state where the relative position degree of freedom forms a bound state corresponding to a localized edge state, i.e. a molecule which is protected by topology.

Eigenstates of the effective Hamiltonian (\ref{Smqheff}) are secular approximation of the exact Floquet states of the system in the moving frame. The latter are eigenstates of the Floquet Hamiltonian which, in the moving frame, reads
\be
H_F=H-(\hat n_1 +\hat n_2)\frac{\omega}{2s}-i\frac{\partial}{\partial t},
\label{SHFnew}
\ee
where $H$ is given in (\ref{Shm}) \cite{Shirley1965}. In order to obtain the exact Floquet states we diagonalize the Floquet Hamiltonian $H_F$ in the Hilbert space extended by the Fourier states $\varphi_k(t)=e^{ik\omega t/s}/\sqrt{T}$ where $T=s2\pi/\omega$, i.e. diagonalize the matrix
\bea
\la n_1',n_2',k'|H_F|n_1,n_2,k\ra&=& \int\limits_0^\pi dx_1 \int\limits_0^\pi dx_2 \int\limits_0^T dt
\cr &\times& \Phi_{n_1',n_2'}^*\varphi_{k'}^*(t) \;H_F\;\Phi_{n_1,n_2}\varphi_k(t).
\cr &&
\label{Shf}
\eea

In Fig.~\ref{Sfig3} a comparison of the spectra obtained by diagonalization of the exact Floquet Hamiltonian (\ref{Shf}) and the effective Hamiltonian (\ref{Smqheff}) is presented. If $\omega$ is sufficiently large, we obtain perfect agreement. It means that the phenomena that we can predict with the help of $H_{\rm eff}$ are reproduced in the exact description of the system.

\begin{figure}[t]
\includegraphics[width=.95\columnwidth]{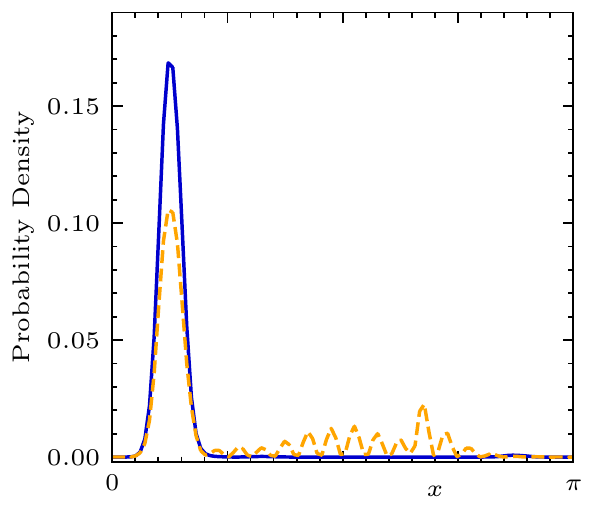}
\caption{Similar plot as in Fig.~3(a) in the Letter but only one edge state is presented (blue line) and $\omega=10^3$. Orange dashed line shows the result of time evolution of the initial Gaussian state (prepared at $\omega t_0/s=\pi/2$, see text) at $\omega t/s=200\pi$ --- the presence of a strong peak indicates the formation of the topological molecule. The units (\ref{Sunits}) are used.}
\label{SfigExp}
\end{figure}

Topological molecules can be realized in ultra-cold atoms. To give a flavor of the experimental parameters, let us consider the following example. Assume that two $^{39}$K atoms in the two different hyperfine states, $|F=1,m_F=0\rangle$ and $|F=1,m_F=-1\rangle$, are trapped in the quasi-1D box potential \cite{Gaunt2013} of longitudinal size of $R=400~\mu$m. The strong transverse confinement is realized by means of the harmonic potential with the frequency $\omega_\perp=2\pi \times 10$~kHz. The atoms are initially prepared in the Gaussian wavepackets of the width $\sigma=58.5~\mu$m located on the opposite sides of the quasi-1D box potential (i.e., at $x=30\;\mu$m and $x=370\;\mu$m) and with the average velocities $\pm 8.0$~mm/s. To observe one of the edge states corresponding to $F_1/F_2=2$ in Fig.~\ref{Sfig3}, the frequency $\omega$ and the maximal amplitude of the inter-species s-wave scattering length modulation [cf. Eq.~(\ref{Sg})] have to be $\omega=2\pi \times 16$~Hz and $3.6$~nm, respectively. This modulation can be achieved by periodic changes of an external magnetic field in time in the vicinity of the Feshbach Resonance at 113.76~G \cite{Tanzi2018}. If the Gaussian wavepackets are prepared at $\omega t_0/s=\pi/2$, then their squared overlap with one of the edge states is 0.4 and in the course of time evolution of the system one will observe signatures of the formation of the topological molecule, see Fig.~\ref{SfigExp}. Note that the intial product of the Gaussian states (which is the easiest to prepare experimentally) belongs both to the symmetric and antisymmetric subspaces, cf. Eqs.~(\ref{asym_subspace})-(\ref{sym_subspace}). Consequently, its maximal possible squared overlap with the edge state is 0.5 and the squared overlap of 0.4 in our example is close to this maximal value. The part of the initial state that belongs to the antisymmetric subspace evolves like for non-interacting particles and consequently it spreads over the entire box and forms a background of the probability density shown in Fig.~\ref{SfigExp}. On this background, a localized peak is clearly visible which is the signature of the topological molecule.

\end{document}